# Energy, linear momentum, and spin and orbital angular momenta of circularly polarized Laguerre-Gaussian wave-packets

Masud Mansuripur
James C. Wyant College of Optical Sciences, The University of Arizona, Tucson, Arizona 85721

**Abstract**. We derive expressions for the energy, linear momentum, and angular momentum content of circularly polarized Laguerre-Gaussian wave-packets propagating in free space. The vectorial nature of the electromagnetic field is taken into account, and the various consequences of paraxial approximation, which is typically invoked in theoretical treatments of the Laguerre-Gaussian beams, are examined.

**1. Introduction**. The Laguerre-Gaussian beams, constructed from Laguerre polynomials with a Gaussian envelope, are an orthogonal set of electromagnetic (EM) field modes that can carry orbital as well as spin angular momentum.[1-4] The phase vorticity of these beams, which can be any arbitrary integer-multiple of $2\pi$ around the axis of propagation, is responsible for their orbital angular momentum, while their degree of circular polarization, which may fall anywhere between the states of left- and right-circular polarization, endows them with a spin angular momentum.[5] A Laguerre-Gaussian beam propagating in a transparent, linear, homogeneous and isotropic medium preserves its general shape, although its overall size and curvature phase-factor change continuously along the axis of propagation. In general, the energy and linear momentum as well as the spin and orbital angular momenta of any wave-packet propagating within a transparent, linear, homogeneous, and isotropic medium are conserved.

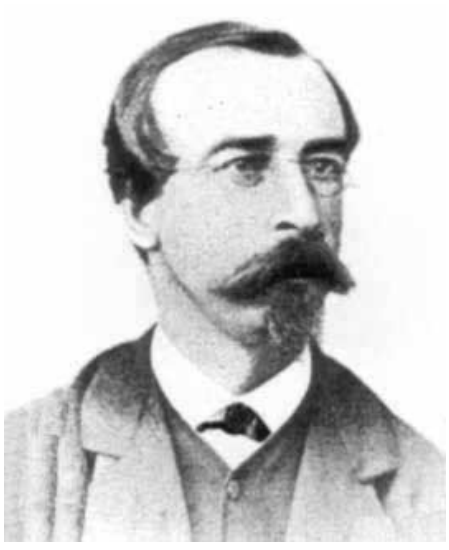

*Edmond Laguerre*
*(1834-1886)*

The goal of the present paper is to provide a framework for analyzing the properties of Laguerre-Gaussian and similar wave-packets while accounting for the vectorial nature of the underlying EM field. We also retain the obliquity factors that are often ignored (or discounted) in tranditional scalar and paraxial treatments of such beams, so that, when necessary, one can obtain quantitative estimates of the effects of paraxial approximation that is part and parcel of the traditional treatment.

We begin in Sec. 2 with the general definition of Laguerre-Gaussian beams as scalar field amplitude distributions in the $xy$-plane of a Cartesian coordinate system. Then, in Sec. 3, we derive the Fourier transform of the beam profile, which is the starting point for all discussions of the physical properties of Laguerre-Gaussian beams. The plane-wave spectrum obtained via spatial Fourier transformation is used in Sec. 4 to arrive at an overall spatio-temporal description of a wave-packet as the basis for our subsequent analysis. Section 5 is devoted to a discussion of the EM energy content of Laguerre-Gaussian beams. Following a brief description of the linear momentum of the wave-packet in Sec. 6, we analyze the wave-packet's angular momentum in Sec. 7, where we derive expressions for both the spin and orbital angular momenta of the packet. The paper closes with a discussion of Bessel-Gaussian beams,[6] and a review of the similarities and differences between this class and the class of Laguerre-Gaussian beams. Straightforward (but often tedious) computational details are relegated to appendices at the end of the paper.

**2. Preliminaries**. Denoting by $(r,\varphi)$ the polar coordinates of the point $(x,y)$, the complex amplitude of a scalar Laguerre-Gaussian beam in the $xy$-plane at $z=0$ is

$$a(r,\varphi) = a_0 r^{\nu} L_n^{\nu}(\alpha r^2) e^{-\beta r^2} e^{\pm \mathrm{i}\nu\varphi}, \tag{1}$$

where $n \geq 0$ and $\nu \geq 0$ are integers, and $a_0, \alpha, \beta$ are arbitrary parameters. The $\pm$ sign is meant to indicate that the phase vorticity is either clockwise or counterclockwise. Here, we have omitted the time-dependence factor $e^{-\mathrm{i}\omega t}$, which is associated with a monochromatic beam of single frequency $\omega$; this factor will make its appearance later on, when we treat wave-packets that contain a narrow band of frequencies.

Relevant properties of the Laguerre polynomials and the Laguerre-Gauss functions are listed in Appendix A. To ensure the orthogonality of profiles that have the same $\nu$ but different $n$ indices, it is necessary to set $\alpha = 2\beta$. The integrated intensity in the $xy$-plane is then given by

$$\int_{r=0}^{\infty} |a_0|^2 r^{2\nu} [L_n^\nu(\alpha r^2)]^2 e^{-\alpha r^2} 2\pi r \mathrm{d}r = \frac{\pi |a_0|^2}{\alpha^{\nu+1}} \int_0^\infty x^\nu [L_n^\nu(x)]^2 e^{-x} \mathrm{d}x = \frac{\pi |a_0|^2 (n+\nu)!}{\alpha^{\nu+1}\, n!}. \tag{2}$$

It is standard practice to introduce a waist parameter $w_0$ by setting $\alpha = 2\beta = 2/w_0^2$, and to normalize the integrated intensity to 1.0 by setting $a_0 = \sqrt{2^{\nu+1}(n!)}/w_0^{\nu+1}\sqrt{\pi(n+\nu)!}$, in which case the complex amplitude profile of the normalized Laguerre-Gauss beam at its waist will be

$$a(x, y, z = 0) = a(r, \varphi, z = 0) = \frac{\sqrt{2(n!)}}{w_0\sqrt{\pi(n+\nu)!}} \left(\frac{\sqrt{2}r}{w_0}\right)^\nu L_n^\nu\left(\frac{2r^2}{w_0^2}\right) e^{-(r/w_0)^2} e^{\pm \mathrm{i}\nu\varphi}. \tag{3}$$

Appendix B shows that monochromatic Laguerre-Gauss beams retain their general shape and overall mathematical structure as they propagate in free space—indeed in any transparent, linear, homogeneous and isotropic medium—provided that one limits the analysis to the paraxial regime,[7,8] where the spatial frequencies are relatively small, and where the vectorial nature of the EM field is of little consequence. This important characteristic of the Laguerre-Gaussian beams, however, is not relevant to the concerns of the present paper, where the focus will be on the angular momentum content of wave-packets formed by the superposition of a narrow band of temporal frequencies $\omega$, all having the same Laguerre-Gaussian structure at their common waist plane. Therefore, in what follows, we shall use the simpler form $a_0 r^\nu L_n^\nu(\alpha r^2) e^{-\alpha r^2/2} e^{\pm \mathrm{i}\nu\varphi}$ of the monochromatic components of these beams, but explicitly take into account the time-dependence factor $e^{-\mathrm{i}\omega t}$, and allow for the amplitude $a_0$ to be a function of the frequency $\omega$.

**3. Fourier transformation**. Denoting the projection $k_x\hat{\boldsymbol{x}} + k_y\hat{\boldsymbol{y}}$ of the $k$-vector onto the $xy$-plane by $(k_\perp, \psi)$ in polar coordinate representation, the Fourier transform of the Laguerre-Gaussian function is found to be

$$\int_{r=0}^{\infty}\int_{\varphi=0}^{2\pi} r^\nu L_n^\nu(\alpha r^2) e^{-\alpha r^2/2} e^{\pm \mathrm{i}\nu\varphi} e^{-\mathrm{i}k_\perp r\cos(\psi-\varphi)} r \mathrm{d}\varphi \mathrm{d}r$$

$$= 2\pi e^{\mathrm{i}\nu(\pm\psi - ½\pi)} \int_0^\infty r^{\nu+1} L_n^\nu(\alpha r^2) e^{-\alpha r^2/2} J_\nu(k_\perp r) \mathrm{d}r$$

$$= (-1)^n 2\pi \alpha^{-(\nu+1)} k_\perp^\nu L_n^\nu(k_\perp^2/\alpha) e^{-k_\perp^2/2\alpha} e^{\mathrm{i}\nu(\pm\psi - ½\pi)}. \tag{4}$$

In addition to Eq.(A5) of Appendix A, the following identities have been used in this derivation:

$$k_x x + k_y y = k_\perp r\cos(\psi - \varphi). \tag{5}$$

$$\int_{-\pi}^{\pi} \exp(-\mathrm{i}\nu\theta + \mathrm{i}z\sin\theta)\, \mathrm{d}\theta = 2\pi J_\nu(z). \qquad (\nu \text{ integer; G\&R 8.411-1})^9 \tag{6}$$

The inverse Fourier transform relation is readily confirmed, as follows:

$$\frac{1}{(2\pi)^2}\int_{k_\perp=0}^{\infty}\int_{\psi=0}^{2\pi} (-1)^n 2\pi\alpha^{-(\nu+1)} k_\perp^\nu L_n^\nu(k_\perp^2/\alpha) e^{-k_\perp^2/2\alpha} e^{\mathrm{i}\nu(\pm\psi-½\pi)}\, e^{\mathrm{i}k_\perp r\cos(\varphi-\psi)} k_\perp \mathrm{d}\psi \mathrm{d}k_\perp$$

$$= (-1)^n \alpha^{-(\nu+1)} e^{\pm \mathrm{i}\nu\varphi} \int_{k_\perp=0}^{\infty} k_\perp^{\nu+1} L_n^\nu(k_\perp^2/\alpha) e^{-k_\perp^2/2\alpha} J_\nu(k_\perp r) \mathrm{d}k_\perp$$

$$= r^\nu L_n^\nu(\alpha r^2) e^{-\alpha r^2/2} e^{\pm \mathrm{i}\nu\varphi}. \tag{7}$$

This decomposition of the Laguerre-Gauss profile into its spatial frequency components is the starting point of our vectorial treatment of the corresponding wave-packet in the next section. It is also the basis for the scalar treatment of monochromatic beams in Appendix B, where each constituent plane-wave is allowed to propagate independently of all the others, before these plane-waves are recombined to form the final amplitude distribution within a destination plane. The dispersion relation in free space, namely, $\boldsymbol{k}\cdot\boldsymbol{k} = k_x^2 + k_y^2 + k_z^2 = k_\perp^2 + k_z^2 = k_0^2$, where $k_0 = \omega/c$ is the wave-number and $c$ is the speed of light in vacuum, shall be used throughout.[10,11]

**4. The Laguerre-Gaussian wave-packet**. Let a thin plate with a Laguerre-Gauss phase-amplitude profile located in the $xy$-plane at $z = 0$ be illuminated by a circularly-polarized plane-wave of frequency $\omega_0$ that is modulated by a fairly broad envelope. Writing $s_\omega$ as an abbreviation for $\mathrm{sign}(\omega)$, and ignoring the $z$-component of the $E$-field for the time being, the distribution emerging from the plate at $z = 0^+$ will be

$$\boldsymbol{E}(x, y, z = 0^+, t) = (2\pi)^{-1} \int_{-\infty}^{\infty} r^\nu L_n^\nu(\alpha r^2) e^{-\alpha r^2/2} e^{\pm \mathrm{i}s_\omega \nu\varphi} E_0(\omega)(\hat{\boldsymbol{x}} + \mathrm{i}s_\omega \hat{\boldsymbol{y}}) e^{-\mathrm{i}\omega t} \mathrm{d}\omega. \tag{8}$$

Aside from the implicit assumption $E_0(-\omega) = E_0^*(\omega)$, note that $s_\omega$ has been introduced in Eq.(8) to ensure that the integrand is Hermitian. Next, we introduce the $z$-component $k_z$ of the $k$-vector, and use the fact that $\boldsymbol{k}\cdot\boldsymbol{E} = 0$ to relate $E_z$ to $E_x$ and $E_y$, as follows:

$$k_z = (\omega/c)\sqrt{1 - (c/\omega)^2(k_x^2 + k_y^2)} = (\omega/c)\sqrt{1 - (ck_\perp/\omega)^2}. \tag{9a}$$

$$\partial k_z/\partial\omega = \left[c\sqrt{1 - (ck_\perp/\omega)^2}\right]^{-1} = (c^2 k_z/\omega)^{-1}. \tag{9b}$$

$$\boldsymbol{E}_0(\boldsymbol{k}, \omega) = E_0(\omega)\left[\hat{\boldsymbol{x}} + \mathrm{i}s_\omega \hat{\boldsymbol{y}} - (k_x + \mathrm{i}s_\omega k_y)\hat{\boldsymbol{z}}/k_z\right]. \tag{10}$$

Invoking the inverse Fourier transform identity given in Eq.(7), we will have

$$\boldsymbol{E}(\boldsymbol{r}, t) = \frac{(-1)^n}{4\pi^2\alpha^{\nu+1}} \iiint_{-\infty}^{\infty} \boldsymbol{E}_0(\boldsymbol{k}, \omega) k_\perp^\nu L_n^\nu(k_\perp^2/\alpha) e^{-k_\perp^2/2\alpha} e^{\mathrm{i}s_\omega \nu(\pm\psi - \frac{1}{2}\pi)} e^{\mathrm{i}(k_x x + k_y y + k_z z - \omega t)} \mathrm{d}k_x \mathrm{d}k_y \mathrm{d}\omega. \tag{11}$$

The corresponding $H$-field is found from Maxwell's equation $\boldsymbol{\nabla} \times \boldsymbol{E} = -\partial_t \boldsymbol{B}$, as follows:

$$\boldsymbol{k} \times \boldsymbol{E}_0 = \mu_0 \omega \boldsymbol{H}_0 \quad \rightarrow \quad \boldsymbol{H}_0 = (c\boldsymbol{k}/\omega) \times \boldsymbol{E}_0/Z_0. \tag{12}$$

$Z_0 = (\mu_0/\varepsilon_0)^{½}$ is the impedance of free space.

$$\boldsymbol{H}_0(\boldsymbol{k}, \omega) = -\frac{cE_0(\omega)}{\omega Z_0}\left\{\left[\frac{k_y(k_x + \mathrm{i}s_\omega k_y)}{k_z} + \mathrm{i}s_\omega k_z\right]\hat{\boldsymbol{x}} - \left[\frac{k_x(k_x + \mathrm{i}s_\omega k_y)}{k_z} + k_z\right]\hat{\boldsymbol{y}} + (k_y - \mathrm{i}s_\omega k_x)\hat{\boldsymbol{z}}\right\}. \tag{13}$$

These equations now yield the following special form of the Poynting vector that will be needed in the next section:

$$\boldsymbol{E}_0(\boldsymbol{k}, \omega) \times \boldsymbol{H}_0(-\boldsymbol{k}, -\omega) = [\boldsymbol{E}_0(\boldsymbol{k}, \omega) \cdot \boldsymbol{E}_0(-\boldsymbol{k}, -\omega)](c\boldsymbol{k}/Z_0\omega) - (c/Z_0\omega)[\underbrace{\boldsymbol{k} \cdot \boldsymbol{E}_0(\boldsymbol{k}, \omega)}_{0}]\boldsymbol{E}_0(-\boldsymbol{k}, -\omega)$$

$$= Z_0^{-1}|E_0(\omega)|^2\left[1 + 1 + (k_x^2 + k_y^2)/k_z^2\right](c\boldsymbol{k}/\omega)$$

$$= Z_0^{-1}|E_0(\omega)|^2[1 + (ck_z/\omega)^{-2}](c\boldsymbol{k}/\omega). \tag{14}$$

**5. Energy content of the wave-packet**. Upon integrating the Poynting vector $\boldsymbol{S}(\boldsymbol{r}, t) = \boldsymbol{E}(\boldsymbol{r}, t) \times \boldsymbol{H}(\boldsymbol{r}, t)$ over an $xy$-plane at an arbitrary fixed point $z = z_0$, and also over all time $t$ (from $-\infty$ to $\infty$), we encounter the following integral whose expression in terms of Dirac's $\delta$-functions considerably simplifies our subsequent calculations:

$$\iiint_{-\infty}^{\infty} \exp[\mathrm{i}(k_x + k_x')x + \mathrm{i}(k_y + k_y')\,y - \mathrm{i}(\omega + \omega')t]\,\mathrm{d}x\mathrm{d}y\mathrm{d}t = (2\pi)^3\delta(k_x + k_x')\delta(k_y + k_y')\delta(\omega + \omega'). \quad (15)$$

Appendix C describes the sifting property of $\delta(\zeta)$ and its derivative $\dot{\delta}(\zeta)$ in some detail. The over-dot represents differentiation with respect to the argument of the function. (Although the discussion in the present section relies solely on the sifting property of the $\delta$-function, the corresponding property of its derivative, $\dot{\delta}$, will be needed in Sec.7.) The sifting property enables one to write $\int_{-\infty}^{\infty}\delta(\zeta + \zeta')f(\zeta')\mathrm{d}\zeta' = f(-\zeta)$ and $\int_{-\infty}^{\infty}\dot{\delta}(\zeta + \zeta')f(\zeta')\mathrm{d}\zeta' = -\dot{f}(-\zeta)$.

With the aid of Eqs.(11), (14), and (15), the total energy content of the Laguerre-Gaussian wave-packet is found to be

$$\begin{aligned}\mathcal{E} &= \iiint_{-\infty}^{\infty} S_z(x, y, z = z_0, t)\mathrm{d}x\mathrm{d}y\mathrm{d}t \\ &= \frac{(-1)^{2n}}{2\pi\alpha^{2(\nu+1)}Z_0}\iiint_{-\infty}^{\infty}|E_0(\omega)|^2[\overbrace{(ck_z/\omega) + (\omega/ck_z)}^{\cong 2}]k_\perp^{2\nu}[L_n^\nu(k_\perp^2/\alpha)]^2 e^{-k_\perp^2/\alpha}\mathrm{d}k_x\mathrm{d}k_y\mathrm{d}\omega \\ &\cong \frac{1}{\pi\alpha^{2(\nu+1)}Z_0}\left[\int_{-\infty}^{\infty}|E_0(\omega)|^2\mathrm{d}\omega\right]\int_0^{\infty}k_\perp^{2\nu}[L_n^\nu(k_\perp^2/\alpha)]^2 e^{-k_\perp^2/\alpha}2\pi k_\perp\mathrm{d}k_\perp \\ &= \frac{1}{Z_0\alpha^{2(\nu+1)}}\left[\int_{-\infty}^{\infty}|E_0(\omega)|^2\mathrm{d}\omega\right]\int_0^{\infty}x^\nu[L_n^\nu(x/\alpha)]^2 e^{-x/\alpha}\mathrm{d}x. \end{aligned} \quad (16)$$

Changing the integration variable to $y = x/\alpha$, invoking Eq.(A4) of Appendix A, and using Parseval's theorem, namely, $\int_{-\infty}^{\infty}|E_0(\omega)|^2\mathrm{d}\omega = 2\pi\int_{-\infty}^{\infty}E^2(t)\mathrm{d}t$, we find

$$\mathcal{E} \cong \frac{2\pi\,(n+\nu)!}{Z_0\alpha^{\nu+1}\,n!}\int_{-\infty}^{\infty}E^2(t)\mathrm{d}t. \quad (17)$$

Compared with Eq.(2), the extra factor of 2 appearing in the above equation is due to the circular polarization of the wave-packet. Paraxial approximation is seen to have been invoked in Eq.(16), when the coefficient $(ck_z/\omega) + (\omega/ck_z)$ of the integrand is approximated with 2. Considering that this coefficient is the sum of the obliquity factor $ck_z/\omega$ and its inverse, the approximation is quite reasonable even when $ck_z/\omega$ drops to as low a value as $\sim 0.75$.

**6. Linear momentum**. The EM momentum density in free space is $\boldsymbol{S}(\boldsymbol{r}, t)/c^2$.[10,11] The circular symmetry of $\boldsymbol{k}$ ensures that the integrated momentum is aligned with the $z$-axis. To find the overall momentum $\boldsymbol{p}$ of a Laguerre-Gauss wave-packet, the integral in Eq.(16) must be taken over $xyz$ while $t$ is fixed at an arbitrary $t_0$. Thus, in place of $\delta(\omega + \omega')$, we end up with $\delta(k_z + k_z') = c(ck_z/\omega)\delta(\omega + \omega')$; see Eq.(9). The integrand in Eq.(16) will now be multiplied by the obliquity factor $ck_z/\omega$. Aside from this small reduction in the overall momentum (caused by the obliquity of $\boldsymbol{k}$), the total momentum of the wave-packet is found to be $\boldsymbol{p} = (\mathcal{E}/c)\hat{\boldsymbol{z}}$. In the language of quantum optics, where the wave packet contains a large number of photons of energy $\hbar\omega$, the linear momentum per photon is approximately $(\hbar\omega/c)\hat{\boldsymbol{z}}$.

We mention in passing that it is also possible to integrate the momentum density $\boldsymbol{S}(\boldsymbol{r}, t)/c^2$ over $x, y, t$ at a fixed plane $z = z_0$. One must take care, however, to account for the rate of flow (per unit time) of linear momentum-density across the fixed plane, which is $c$ times $ck_z/\omega$, due to the obliquity factor associated with each constituent plane-wave. Once this obliquity factor is taken into account, the result of integration over $x, y, t$ at fixed $z = z_0$ will be identical with that obtained by integration over $x, y, z$ at fixed $t = t_0$.

**7. Angular momentum**. We compute the total angular momentum of our circularly polarized wave-packet, whose $E$-field distribution is given by Eq.(11), with $\boldsymbol{E}_0(\boldsymbol{k},\omega)$ given in Eq.(10). The expression for the $H$-field is similar to that of the $E$-field in Eq.(11), except that $\boldsymbol{E}_0(\boldsymbol{k},\omega)$ is now replaced with $\boldsymbol{H}_0(\boldsymbol{k},\omega)$ of Eq.(13). Integrating the EM angular momentum density of the packet over the entire $xyz$ space at a fixed (but otherwise arbitrary) instant in time, we will have[10,11]

$$\boldsymbol{\mathcal{L}}(t_0) = c^{-2}\iiint_{-\infty}^{\infty}\boldsymbol{r}\times\boldsymbol{S}(\boldsymbol{r},t=t_0)\mathrm{d}x\mathrm{d}y\mathrm{d}z. \tag{18}$$

Upon substituting $\boldsymbol{E}\times\boldsymbol{H}$ for the Poynting vector $\boldsymbol{S}$, and rearranging the order of integration, Eq.(18) becomes

$\boxed{E_0(k',\omega')k_\perp'^{\nu}L_n^{\nu}(k_\perp'^2/\alpha)e^{-k_\perp'^2/2\alpha}e^{\mathrm{i}s'_\omega\nu(\pm\psi'-½\pi)}}$ ↓ $\quad$ $\boxed{H_0(k,\omega)k_\perp^{\nu}L_n^{\nu}(k_\perp^2/\alpha)e^{-k_\perp^2/2\alpha}e^{\mathrm{i}s_\omega\nu(\pm\psi-½\pi)}}$ ↓

$$\boldsymbol{\mathcal{L}} = \frac{(-1)^{2n}}{(2\pi)^4\alpha^{2(\nu+1)}c^2}\int_{-\infty}^{\infty}\boldsymbol{r}\times\left[\int_{-\infty}^{\infty}\boldsymbol{E}(\boldsymbol{k}',\omega')e^{\mathrm{i}(\boldsymbol{k}'\cdot\boldsymbol{r}-\omega' t_0)}\mathrm{d}k'_x\mathrm{d}k'_y\mathrm{d}\omega'\times\int_{-\infty}^{\infty}\boldsymbol{H}(\boldsymbol{k},\omega)e^{\mathrm{i}(\boldsymbol{k}\cdot\boldsymbol{r}-\omega t_0)}\mathrm{d}k_x\mathrm{d}k_y\mathrm{d}\omega\right]\mathrm{d}x\mathrm{d}y\mathrm{d}z$$

$$= \frac{1}{(2\pi)^4\alpha^{2(\nu+1)}c^2}\int_{-\infty}^{\infty}\left\{\int_{-\infty}^{\infty}(x\hat{\boldsymbol{x}}+y\hat{\boldsymbol{y}}+z\hat{\boldsymbol{z}})\exp[\mathrm{i}(\boldsymbol{k}+\boldsymbol{k}')\cdot(x\hat{\boldsymbol{x}}+y\hat{\boldsymbol{y}}+z\hat{\boldsymbol{z}})]\,\mathrm{d}x\mathrm{d}y\mathrm{d}z\right\}$$

$$\times[\boldsymbol{E}(\boldsymbol{k}',\omega')\times\boldsymbol{H}(\boldsymbol{k},\omega)]\exp[-\mathrm{i}(\omega+\omega')t_0]\,\mathrm{d}k'_x\mathrm{d}k'_y\mathrm{d}\omega'\mathrm{d}k_x\mathrm{d}k_y\mathrm{d}\omega. \tag{19}$$

In what follows, we set $t_0 = 0$ to simplify the algebra, although the same line of reasoning can be used to show that the end result is independent of the value of $t_0$. (In other words, the wave-packet's total angular momentum is conserved as it propagates through free space.) In Eq.(19), the inner integral over $xyz$ contains three terms of the following general form:

$$\iiint_{-\infty}^{\infty}x\exp[\mathrm{i}(k_x+k'_x)x+\mathrm{i}(k_y+k'_y)y+\mathrm{i}(k_z+k'_z)z]\,\mathrm{d}x\mathrm{d}y\mathrm{d}z = -\mathrm{i}(2\pi)^3\dot{\delta}(k_x+k'_x)\delta(k_y+k'_y)\delta(k_z+k'_z). \tag{20}$$

Thus, upon carrying out the inner integral, Eq.(19) becomes

$$\boldsymbol{\mathcal{L}}(0) = \frac{1}{\mathrm{i}2\pi c^2\alpha^{2(\nu+1)}}\int_{-\infty}^{\infty}[\dot{\delta}(k_x+k'_x)\delta(k_y+k'_y)\delta(k_z+k'_z)\hat{\boldsymbol{x}}+\delta(k_x+k'_x)\dot{\delta}(k_y+k'_y)\delta(k_z+k'_z)\hat{\boldsymbol{y}}$$

$$+\delta(k_x+k'_x)\delta(k_y+k'_y)\dot{\delta}(k_z+k'_z)\hat{\boldsymbol{z}}]\times[\boldsymbol{E}(\boldsymbol{k}',\omega')\times\boldsymbol{H}(\boldsymbol{k},\omega)]\mathrm{d}k'_x\mathrm{d}k'_y\mathrm{d}\omega'\mathrm{d}k_x\mathrm{d}k_y\mathrm{d}\omega. \tag{21}$$

The six-dimensional integral in Eq.(21) is amenable to further simplifications. Observing that

$$\boldsymbol{r}\times(\boldsymbol{E}\times\boldsymbol{H}) = (x\hat{\boldsymbol{x}}+y\hat{\boldsymbol{y}}+z\hat{\boldsymbol{z}})\times[(E_yH_z-E_zH_y)\hat{\boldsymbol{x}}+(E_zH_x-E_xH_z)\hat{\boldsymbol{y}}+(E_xH_y-E_yH_x)\hat{\boldsymbol{z}}]$$

$$= [yE_xH_y+zE_xH_z-(yE_y+zE_z)H_x]\hat{\boldsymbol{x}}+[zE_yH_z+xE_yH_x-(zE_z+xE_x)H_y]\hat{\boldsymbol{y}}$$

$$+[xE_zH_x+yE_zH_y-(xE_x+yE_y)H_z]\hat{\boldsymbol{z}}, \tag{22}$$

we write $\boldsymbol{E}(\boldsymbol{k}',\omega')$ and $\boldsymbol{H}(\boldsymbol{k},\omega)$ as $\boldsymbol{E}'$ and $\boldsymbol{H}$, and abbreviate $\dot{\delta}(k_x+k'_x)\delta(k_y+k'_y)\delta(k_z+k'_z)$ as $\dot{\delta}_x\delta_y\delta_z$, etc., to arrive at the following streamlined version of Eq.(21):

$$\mathcal{L}_x = \frac{1}{\mathrm{i}2\pi c^2\alpha^{2(\nu+1)}}\int_{-\infty}^{\infty}[\delta_x\dot{\delta}_y\delta_zE'_xH_y+\delta_x\delta_y\dot{\delta}_zE'_xH_z-(\delta_x\dot{\delta}_y\delta_zE'_y+\delta_x\delta_y\dot{\delta}_zE'_z)H_x]\mathrm{d}k'_x\mathrm{d}k'_y\mathrm{d}\omega'\mathrm{d}k_x\mathrm{d}k_y\mathrm{d}\omega. \tag{23a}$$

$$\mathcal{L}_y = \frac{1}{\mathrm{i}2\pi c^2\alpha^{2(\nu+1)}}\int_{-\infty}^{\infty}[\delta_x\delta_y\dot{\delta}_zE'_yH_z+\dot{\delta}_x\delta_y\delta_zE'_yH_x-(\delta_x\delta_y\dot{\delta}_zE'_z+\dot{\delta}_x\delta_y\delta_zE'_x)H_y]\mathrm{d}k'_x\mathrm{d}k'_y\mathrm{d}\omega'\mathrm{d}k_x\mathrm{d}k_y\mathrm{d}\omega. \tag{23b}$$

$$\mathcal{L}_z = \frac{1}{\mathrm{i}2\pi c^2\alpha^{2(\nu+1)}}\int_{-\infty}^{\infty}[\dot{\delta}_x\delta_y\delta_zE'_zH_x+\delta_x\dot{\delta}_y\delta_zE'_zH_y-(\dot{\delta}_x\delta_y\delta_zE'_x+\delta_x\dot{\delta}_y\delta_zE'_y)H_z]\mathrm{d}k'_x\mathrm{d}k'_y\mathrm{d}\omega'\mathrm{d}k_x\mathrm{d}k_y\mathrm{d}\omega. \tag{23c}$$

In these equations, $\dot{\delta}_z = \dot{\delta}(k_z+k'_z)$ is always accompanied by $\delta_x\delta_y$, which means that, upon integrating over $k'_x$ and $k'_y$, the argument of $\dot{\delta}(k_z+k'_z)$ reduces to a function of $k_x$, $k_y$, $\omega$ and $\omega'$, in which case, in accordance with Eqs.(9) and Appendix C, $\delta_x\delta_y\dot{\delta}_z$ can be written as

$$\delta_x\delta_y\dot{\delta}_z = (c^2k_z/\omega)^2\delta(k_x+k'_x)\delta(k_y+k'_y)\dot{\delta}(\omega+\omega'). \quad (24)$$

The other $\delta$-function triplets in Eqs.(23) require further attention, as differentiation with respect to either $k'_x$ or $k'_y$ affects $\delta(k_z+k'_z)$ as well. It is not difficult to show that

$$\dot{\delta}_x\delta_y\delta_z = (c^2k_z/\omega)\dot{\delta}(k_x+k'_x)\delta(k_y+k'_y)\delta(\omega+\omega') - (k_x/k'_z)(c^2k_z/\omega)^2\delta(k_x+k'_x)\delta(k_y+k'_y)\dot{\delta}(\omega+\omega'). \quad (25)$$

$$\delta_x\dot{\delta}_y\delta_z = (c^2k_z/\omega)\delta(k_x+k'_x)\dot{\delta}(k_y+k'_y)\delta(\omega+\omega') - (k_y/k'_z)(c^2k_z/\omega)^2\delta(k_x+k'_x)\delta(k_y+k'_y)\dot{\delta}(\omega+\omega'). \quad (26)$$

The integrals in Eqs.(23) are evaluated in Appendix D. The end result is that $\mathcal{L}_x = \mathcal{L}_y = 0$, and that the $z$-component of the wave-packet's total angular momentum is given by

$$\mathcal{L}_z = \frac{1\pm\nu}{Z_0\alpha^{2(\nu+1)}}\int_{\omega=-\infty}^{\infty}\frac{|E_0(\omega)|^2}{|\omega|}\int_{k_\perp=0}^{\infty}\overbrace{[(ck_z/\omega)+(\omega/ck_z)]}^{\cong 2}k_\perp^{2\nu}[L_n^\nu(k_\perp^2/\alpha)]^2e^{-k_\perp^2/\alpha}k_\perp\mathrm{d}k_\perp\mathrm{d}\omega. \quad (27)$$

Assuming the obliquity factor $ck_z/\omega$ is close to 1, the integral over the $k_xk_y$-plane yields $\alpha^{\nu+1}(n+\nu)!/n!$. Also, for a narrowband wave-packet, since $|\omega| \cong \omega_0$, the integral of $|E(\omega)|^2$ over $\omega$ gives $2\pi\int_{-\infty}^{\infty}E^2(t)\mathrm{d}t$. This makes the overall $\mathcal{L}_z$ very nearly equal to $(1\pm\nu)\mathcal{E}/\omega_0$; see Eq.(17). While the contribution of orbital angular momentum to $\mathcal{L}_z$ is $\pm\nu\mathcal{E}/\omega_0$, that of the spin angular momentum (due to circular polarization) is $\mathcal{E}/\omega_0$. Had we chosen the opposite sense of circular polarization at the outset, we would have found $\mathcal{L}_z \cong (-1\pm\nu)\mathcal{E}/\omega_0$. In the language of quantum optics, a wave-packet containing a large number of photons of energy $\hbar\omega_0$, carries orbital and spin angular momenta per photon in the amounts of $\pm\nu\hbar$ and $\pm\hbar$, respectively. It is noteworthy that, while for a given energy content $\mathcal{E}$ the linear momentum of the wave-packet steadily declines with a rising obliquity factor, the packet's angular momentum retains its value as the beam gets further and further away from the paraxial regime.

**8. Concluding remarks**. In this paper, we have described some of the important properties of the Laguerre-Gaussian beams, with emphasis on the vectorial nature of the underlying EM field, and with an eye toward the degree of approximation that the paraxial treatment of wave-packets would inevitably introduce into the quantitative results of our analysis. Similar arguments can be used to evaluate the energy, linear momentum, and angular momentum content of Bessel-Gauss beams in the paraxial regime.[6,12] The scalar amplitude profile of a Bessel-Gauss beam in the $xy$-plane at $z=0$ is defined as

$$a(x,y,z=0) = a(r,\varphi,z=0) = a_0J_\nu(r/r_0)e^{-(r/w_0)^2}e^{\pm\mathrm{i}\nu\varphi}. \quad (28)$$

Here, the amplitude $a_0$ is a (generally complex) constant; $J_\nu(\cdot)$ is a Bessel function of the first kind, integer order $\nu \geq 0$; the parameters $r_0 > 0$ and $w_0 > 0$ are real-valued constants; and the $\pm$ signs specify the sense of the beam's vorticity as either clockwise or counterclockwise. The beam's integrated intensity is found to be (see Appendix E):

$$\int_0^{\infty}|a_0|^2J_\nu^2(r/r_0)e^{-2(r/w_0)^2}2\pi r\mathrm{d}r = \tfrac{1}{2}\pi|a_0|^2w_0^2e^{-(w_0/2r_0)^2}I_\nu(w_0^2/4r_0^2). \quad (29)$$

In the above equation $I_\nu(x) = \mathrm{i}^{-\nu}J_\nu(\mathrm{i}x)$ is a modified Bessel function of the first kind, order $\nu$. One difference between the Laguerre-Gauss and Bessel-Gauss beams is that the latter class does *not* comprise an orthogonal set of basis functions that could be used to expand arbitrary beam profiles as a superposition of the basis functions. Another difference is that, upon propagation through transparent, linear, homogeneous, isotropic media, a Bessel-Gauss beam does *not* preserve its initial shape. Appendix E contains a brief description of the paraxial propagation of Bessel-Gauss beams in free space, where it is shown that the argument $r/r_0$ of $J_\nu(\cdot)$ in Eq.(28) becomes progressively more complex-valued along the propagation direction. Despite these differences, the Bessel-Gauss beams carry the same amount of linear momentum as well as spin and orbital angular momenta per photon as do the Laguerre-Gauss beams. The same approximations pertaining to the paraxial regime apply to both types of beams.

## Appendix A

The relevant formulas for Laguerre polynomials and Laguerre-Gauss functions are listed in Gradshteyn and Ryzhik's *Table of Integrals, Series, and Products*.[9] The Laguerre polynomials $L_n^\nu(x)$ satisfy the following differential equation:

$$x\frac{\mathrm{d}^2 f(x)}{\mathrm{d}x^2} + (\nu - x + 1)\frac{\mathrm{d}f(x)}{\mathrm{d}x} + nf(x) = 0. \qquad \text{(Gradshteyn \& Ryzhik 8.979)} \quad \text{(A1)}$$

Polynomial representation:

$$L_n^\nu(x) = \sum_{m=0}^{n} (-1)^m \binom{n+\nu}{n-m} \frac{x^m}{m!}. \qquad \text{(G\&R 8.970)} \quad \text{(A2)}$$

Rodrigues' formula:

$$L_n^\nu(x) = \frac{1}{n!} e^x x^{-\nu} \frac{\mathrm{d}^n}{\mathrm{d}x^n}\left(e^{-x} x^{n+\nu}\right). \qquad \text{(G\&R 8.970)} \quad \text{(A3)}$$

Orthogonality:

$$\int_0^\infty x^\nu L_n^\nu(x) L_m^\nu(x) e^{-x} \mathrm{d}x = \begin{cases} 0; & m \neq n, \\ \Gamma(\nu+1)\binom{n+\nu}{n}; & m = n. \end{cases} \qquad \text{(G\&R 8.980)} \quad \text{(A4)}$$

Two-dimensional Fourier transformation:

$$\int_0^\infty x^{\nu+1} L_n^\nu(\alpha x^2) e^{-\beta x^2} J_\nu(kx) \mathrm{d}x = \frac{[1-(\alpha/\beta)]^n}{(2\beta)^{\nu+1}} k^\nu L_n^\nu\left[\frac{\alpha k^2}{4\beta(\alpha-\beta)}\right] e^{-k^2/4\beta}. \qquad \text{(G\&R 7.421-4)} \quad \text{(A5)}$$

Functional relations:

$$x\frac{\mathrm{d}}{\mathrm{d}x} L_n^\nu(x) = (n+1)L_{n+1}^\nu(x) - (n+\nu+1-x)L_n^\nu(x). \qquad \text{(G\&R 8.971-3,4)} \quad \text{(A6)}$$

$$x\frac{\mathrm{d}}{\mathrm{d}x} L_n^\nu(x) = nL_n^\nu(x) - (n+\nu)L_{n-1}^\nu(x). \qquad \text{(A7)}$$

$$xL_n^{\nu+1}(x) = (n+\nu+1)L_n^\nu(x) - (n+1)L_{n+1}^\nu(x) = (n+\nu)L_{n-1}^\nu(x) - (n-x)L_n^\nu(x). \qquad \text{(A8)}$$

## Appendix B

In the paraxial approximation,[7,8] a scalar Laguerre-Gaussian beam retains its general form as it propagates through a transparent, linear, homogeneous and isotropic medium. We set $k_z \cong k_0[1 - ½(k_\perp/k_0)^2]$ for paraxial propagation in free space, and $a_0 = \sqrt{2^{\nu+1} n!/\pi(n+\nu)!}/w_0^{\nu+1}$ to normalize the beam's integrated intensity to 1. For a single-frequency Laguerre-Gauss beam (i.e., one whose time-dependence factor is $e^{-\mathrm{i}\omega t}$), Eq.(11) can be written as

$$a(x,y,z) = A_0 \iint_{-\infty}^{\infty} k_\perp^\nu L_n^\nu[(w_0 k_\perp)^2/2] e^{-¼(w_0 k_\perp)^2} e^{\mathrm{i}\nu(\pm\psi - ½\pi)} e^{\mathrm{i}(k_x x + k_y y + k_z z)} \mathrm{d}k_x \mathrm{d}k_y, \tag{B1}$$

where

$$A_0 = \frac{(-1)^n \sqrt{n!}}{2\pi\sqrt{\pi(n+\nu)!}} \left(\frac{w_0}{\sqrt{2}}\right)^{\nu+1}. \tag{B2}$$

The light amplitude distribution in the $xy$-plane at a distance $z$ from the initial distribution is now evaluated as follows:

$$\begin{aligned}
a(x,y,z) &= A_0 \int_{k_\perp=0}^{\infty} k_\perp^{\nu+1} L_n^\nu[(w_0 k_\perp)^2/2] e^{-¼(w_0 k_\perp)^2} e^{\mathrm{i}k_z z} \left[\int_{\psi=0}^{2\pi} e^{\mathrm{i}\nu(\pm\psi-½\pi)} e^{\mathrm{i}k_\perp r\cos(\psi-\varphi)} \mathrm{d}\psi\right] \mathrm{d}k_\perp \\
&\cong 2\pi A_0 e^{\pm\mathrm{i}\nu\varphi} \int_0^\infty k_\perp^{\nu+1} L_n^\nu[(w_0 k_\perp)^2/2] e^{-¼(w_0 k_\perp)^2} e^{\mathrm{i}k_0[1-½(k_\perp/k_0)^2]z} J_\nu(k_\perp r) \mathrm{d}k_\perp \\
&= 2\pi A_0 e^{\pm\mathrm{i}\nu\varphi} e^{\mathrm{i}k_0 z} \int_0^\infty k_\perp^{\nu+1} L_n^\nu[(w_0 k_\perp)^2/2] e^{-¼[w_0^2+\mathrm{i}(2z/k_0)]k_\perp^2} J_\nu(k_\perp r) \mathrm{d}k_\perp \\
&= \frac{\sqrt{2^{\nu+1} n!}\ \exp[-\mathrm{i}(2n+\nu+1)\tan^{-1}(2z/k_0 w_0^2)]}{\sqrt{\pi(n+\nu)!}\ [w_0^2 + (2z/k_0 w_0)^2]^{½}} e^{\mathrm{i}k_0 z} \\
&\quad \times \left[r/\sqrt{w_0^2 + (2z/k_0 w_0)^2}\right]^\nu L_n^\nu\left[\frac{2r^2}{w_0^2 + (2z/k_0 w_0)^2}\right] \exp\left[-\frac{r^2}{w_0^2 + \mathrm{i}(2z/k_0)}\right] e^{\pm\mathrm{i}\nu\varphi} \\
&= \frac{\sqrt{2^{\nu+1} n!}\exp[-\mathrm{i}(2n+\nu+1)\tan^{-1}(2z/k_0 w_0^2)]}{\sqrt{\pi(n+\nu)!}\,[w_0^2+(2z/k_0 w_0)^2]^{½}} e^{\mathrm{i}k_0 z} \exp\left(\frac{\mathrm{i}k_0 r^2}{2z[1+(k_0 w_0^2/2z)^2]}\right) \\
&\quad \times \left[r/\sqrt{w_0^2 + (2z/k_0 w_0)^2}\right]^\nu L_n^\nu\left[\frac{2r^2}{w_0^2 + (2z/k_0 w_0)^2}\right] \exp\left[-\frac{r^2}{w_0^2 + (2z/k_0 w_0)^2}\right] e^{\pm\mathrm{i}\nu\varphi} \\
&= \frac{\sqrt{2^{\nu+1} n!}}{\sqrt{\pi(n+\nu)!}\,[w_0^2+(2z/k_0 w_0)^2]^{½}} \\
&\quad \times \exp\left\{\mathrm{i}\left[k_0 z + \frac{k_0 r^2}{2z[1+(k_0 w_0^2/2z)^2]} - (2n+\nu+1)\tan^{-1}\left(\frac{2z}{k_0 w_0^2}\right)\right]\right\} \\
&\quad \times \left[r/\sqrt{w_0^2 + (2z/k_0 w_0)^2}\right]^\nu L_n^\nu\left[\frac{2r^2}{w_0^2+(2z/k_0 w_0)^2}\right] \exp\left[-\frac{r^2}{w_0^2+(2z/k_0 w_0)^2}\right] e^{\pm\mathrm{i}\nu\varphi}.
\end{aligned} \tag{B3}$$

At a distance $z$ from the initial distribution, one now observes that the beam's waist and its radius of curvature are $w(z) = \sqrt{w_0^2 + (2z/k_0 w_0)^2}$ and $R(z) = z[1 + (k_0 w_0^2/2z)^2]$, respectively. The phase-shift $\chi(z) = (2n + \nu + 1)\tan^{-1}(2z/k_0 w_0^2)$ appearing in Eq.(B3) is commonly referred to as the Gouy phase.[8] The light amplitude distribution in the $xy$-plane at $z$ is thus given by

$$a(x,y,z) \cong \frac{\sqrt{2(n!)}}{\sqrt{\pi(n+\nu)!}\, w(z)} \exp\left\{\mathrm{i}\left[k_0 z + \frac{k_0 r^2}{2R(z)} - \chi(z)\right]\right\} \left[\frac{\sqrt{2}\, r}{w(z)}\right]^\nu L_n^\nu\left[\frac{2r^2}{w^2(z)}\right] \exp\left[-\frac{r^2}{w^2(z)}\right] e^{\pm\mathrm{i}\nu\varphi}. \tag{B4}$$

It is straightforward to verify that the integrated intensity across the $xy$-plane at $z$ is the same as that of the initial distribution at $z = 0$, that is, $\iint_{-\infty}^{\infty} |a(x,y,z)|^2 \mathrm{d}x\mathrm{d}y = 1$.

## Appendix C

The scaled $\delta$-function $\delta(\alpha x)$, where $\alpha \neq 0$ is a real-valued scaling parameter, is equal to $|\alpha|^{-1}\delta(x)$. To see this, note that

$$\int_{-\infty}^{\infty}\delta(\alpha x)f(x)\mathrm{d}x = \alpha^{-1}\int_{\mp\infty}^{\pm\infty}\delta(y)f(y/\alpha)\mathrm{d}y = |\alpha|^{-1}f(0). \quad \text{(C1)}$$

In contrast, the scaled derivative $\dot{\delta}(\alpha x)$ of the $\delta$-function, where the over-dot represents differentiation with respect to the argument of the function, is $\mathrm{sign}(\alpha)\alpha^{-2}\dot{\delta}(x)$. For a straightforward proof, note that

$$\int_{-\infty}^{\infty}\dot{\delta}(\alpha x)f(x)\mathrm{d}x = \alpha^{-1}\int_{\mp\infty}^{\pm\infty}\dot{\delta}(y)f(y/\alpha)\mathrm{d}y = |\alpha|^{-1}\big[-\alpha^{-1}\dot{f}(0)\big] = -\mathrm{sign}(\alpha)|\alpha|^{-2}\dot{f}(0). \quad \text{(C2)}$$

Below is an alternative proof that relies on the method of integration by parts:

$$\int_{-\infty}^{\infty}\dot{\delta}(\alpha x)f(x)\mathrm{d}x = \alpha^{-1}\delta(\alpha x)f(x)|_{-\infty}^{\infty} - \int_{-\infty}^{\infty}\alpha^{-1}\delta(\alpha x)\dot{f}(x)\mathrm{d}x = -\mathrm{sign}(\alpha)\alpha^{-2}\dot{f}(0). \quad \text{(C3)}$$

Note that $\dot{\delta}(\alpha x)$, where the over-dot represents differentiation with respect to the argument of the $\delta$-function, should *not* be confused with the derivative with respect to $x$ of $\delta(\alpha x)$, which is $|\alpha|^{-1}\dot{\delta}(x)$. Below are two different ways of demonstrating the validity of the above statement:

i) $$\frac{\mathrm{d}}{\mathrm{d}x}\delta(\alpha x) = \frac{\mathrm{d}}{\mathrm{d}x}[|\alpha|^{-1}\delta(x)] = |\alpha|^{-1}\dot{\delta}(x). \quad \text{(C4)}$$

ii) $$\frac{\mathrm{d}}{\mathrm{d}x}\delta(\alpha x) = \alpha\dot{\delta}(\alpha x) = \alpha\big[\mathrm{sign}(\alpha)\alpha^{-2}\dot{\delta}(x)\big] = |\alpha|^{-1}\dot{\delta}(x). \quad \text{(C5)}$$

In general, one may have a function such as $\delta[g(x)]$, where $g(x)$ goes to zero at one or more locations along the $x$-axis. Denoting by $x_0$ a typical zero of $g(x)$, one can replace (in the immediate vicinity of $x_0$) the argument of the $\delta$-function by $\alpha(x - x_0)$, where $\alpha = \dot{g}(x_0)$ is the slope of $g(x)$ at $x = x_0$. At each zero of $g(x)$, one may then write

$$\delta[g(x)] = |\dot{g}(x_0)|^{-1}\delta(x - x_0), \quad \text{(C6)}$$

$$\dot{\delta}[g(x)] = \mathrm{sign}[\dot{g}(x_0)] \times |\dot{g}(x_0)|^{-2}\dot{\delta}(x - x_0). \quad \text{(C7)}$$

Needless to say, additional measures must be taken in the special circumstance where $\dot{g}(x_0) = 0$.

## Appendix D

The integrals in Eqs.(23) are evaluated in several steps. We begin by defining the functions $A$ and $B$ of the variable $k_\perp = (k_x^2 + k_y^2)^{½}$ (with $\alpha$ and $\nu$ as parameters), as follows:

$$A = k_\perp^\nu L_n^\nu(k_\perp^2/\alpha)e^{-k_\perp^2/2\alpha} \quad \rightarrow \quad \partial_{k_x}A = Bk_x \quad \text{and} \quad \partial_{k_y}A = Bk_y, \quad \text{where}$$

$$B = [\nu k_\perp^{\nu-2}L_n^\nu(k_\perp^2/\alpha) + \alpha^{-1}k_\perp^\nu L_n^\nu(k_\perp^2/\alpha) - 2\alpha^{-1}k_\perp^\nu L_n^{\nu+1}(k_\perp^2/\alpha)]e^{-k_\perp^2/2\alpha}. \quad \text{(D1)}$$

In the next step, we derive the partial derivatives with respect to $k_x$ and $k_y$ of the angle $\psi$, that is,

$$\psi = \tan^{-1}(k_y/k_x) \quad \rightarrow \quad \partial_{k_x}\psi = -k_y/k_\perp^2 \quad \text{and} \quad \partial_{k_y}\psi = k_x/k_\perp^2. \quad \text{(D2)}$$

The third step involves evaluating triple integrals such as

$$\iiint_{-\infty}^{\infty}\dot{\delta}(k_x + k_x')\delta(k_y + k_y')\delta(\omega + \omega')E_x(\boldsymbol{k}', \omega')k_\perp'^\nu L_n^\nu(k_\perp'^2/\alpha)e^{-k_\perp'^2/2\alpha}e^{\mathrm{i}s_\omega'\nu(\pm\psi' - ½\pi)}\mathrm{d}k_x'\mathrm{d}k_y'\mathrm{d}\omega'. \quad \text{(D3)}$$

Here, one must first differentiate with respect to $k_x'$, then substitute $(-k_x, -k_y, -\omega)$ for $(k_x', k_y', \omega')$. Subsequently, the sign of the entire expression must be reversed. Below, we carry out the differentiation step for each term that appears in Eqs.(23). The substitution and sign reversal will be done afterward. We find

$$\dot{\delta}_x\delta_y\delta_t E'_x = E_0(\omega')\left[Bk'_x - A(\pm \mathrm{i}s'_\omega\nu)\frac{k'_y}{k'^2_\perp}\right] e^{\mathrm{i}s'_\omega\nu(\pm\psi'-\frac{1}{2}\pi)}. \tag{D4}$$

$$\delta_x\dot{\delta}_y\delta_t E'_x = E_0(\omega')\left[Bk'_y + A(\pm \mathrm{i}s'_\omega\nu)\frac{k'_x}{k'^2_\perp}\right] e^{\mathrm{i}s'_\omega\nu(\pm\psi'-\frac{1}{2}\pi)}. \tag{D5}$$

$$\dot{\delta}_x\delta_y\delta_t E'_y = \mathrm{i}s_\omega E_0(\omega')\left[Bk'_x - A(\pm \mathrm{i}s'_\omega\nu)\frac{k'_y}{k'^2_\perp}\right] e^{\mathrm{i}s'_\omega\nu(\pm\psi'-\frac{1}{2}\pi)}. \tag{D6}$$

$$\delta_x\dot{\delta}_y\delta_t E'_y = \mathrm{i}s_\omega E_0(\omega')\left[Bk'_y + A(\pm \mathrm{i}s'_\omega\nu)\frac{k'_x}{k'^2_\perp}\right] e^{\mathrm{i}s'_\omega\nu(\pm\psi'-\frac{1}{2}\pi)}. \tag{D7}$$

$$\dot{\delta}_x\delta_y\delta_t E'_z = -E_0(\omega')\left[\frac{(k'^2_x+k'^2_z)+\mathrm{i}s'_\omega k'_x k'_y}{k'^3_z}A + \frac{(k'_x+\mathrm{i}s'_\omega k'_y)k'_x}{k'_z}B - \frac{(k'_x+\mathrm{i}s'_\omega k'_y)k'_y}{k'_z k'^2_\perp}A(\pm \mathrm{i}s'_\omega\nu)\right] e^{\mathrm{i}s'_\omega\nu(\pm\psi'-\frac{1}{2}\pi)}. \tag{D8}$$

$$\delta_x\dot{\delta}_y\delta_t E'_z = -E_0(\omega')\left[\frac{\mathrm{i}s'_\omega(k'^2_y+k'^2_z)+k'_x k'_y}{k'^3_z}A + \frac{(k'_x+\mathrm{i}s'_\omega k'_y)k'_y}{k'_z}B + \frac{(k'_x+\mathrm{i}s'_\omega k'_y)k'_x}{k'_z k'^2_\perp}A(\pm \mathrm{i}s'_\omega\nu)\right] e^{\mathrm{i}s'_\omega\nu(\pm\psi'-\frac{1}{2}\pi)}. \tag{D9}$$

Replacing $(k'_x, k'_y, \omega')$ with $(-k_x, -k_y, -\omega)$ and flipping the signs of the resulting expressions, we arrive at

$$\dot{\delta}_x\delta_y\delta_t E_x = -E_0(-\omega)\left[-Bk_x - A(\pm \mathrm{i}s_\omega\nu)\frac{k_y}{k^2_\perp}\right] e^{-\mathrm{i}s_\omega\nu(\pm\psi-\frac{1}{2}\pi)}. \tag{D10}$$

$$\delta_x\dot{\delta}_y\delta_t E_x = -E_0(-\omega)\left[-Bk_y + A(\pm \mathrm{i}s_\omega\nu)\frac{k_x}{k^2_\perp}\right] e^{-\mathrm{i}s_\omega\nu(\pm\psi-\frac{1}{2}\pi)}. \tag{D11}$$

$\psi$ changes by $\pi$, resulting in a factor of $(-1)^\nu$, which cancels a similar factor in $A$ and $B$ due to $k_\perp^\nu$ and $k_\perp^{\nu-2}$.

$$\dot{\delta}_x\delta_y\delta_t E_y = -\mathrm{i}s_\omega E_0(-\omega)\left[Bk_x + A(\pm \mathrm{i}s_\omega\nu)\frac{k_y}{k^2_\perp}\right] e^{-\mathrm{i}s_\omega\nu(\pm\psi-\frac{1}{2}\pi)}. \tag{D12}$$

$$\delta_x\dot{\delta}_y\delta_t E_y = -\mathrm{i}s_\omega E_0(-\omega)\left[Bk_y - A(\pm \mathrm{i}s_\omega\nu)\frac{k_x}{k^2_\perp}\right] e^{-\mathrm{i}s_\omega\nu(\pm\psi-\frac{1}{2}\pi)}. \tag{D13}$$

$$\dot{\delta}_x\delta_y\delta_t E_z = -E_0(-\omega)\left[\frac{(k^2_x+k^2_z)-\mathrm{i}s_\omega k_x k_y}{k^3_z}A + \frac{(k_x-\mathrm{i}s_\omega k_y)k_x}{k_z}B + \frac{(k_x-\mathrm{i}s_\omega k_y)k_y}{k_z k^2_\perp}A(\pm \mathrm{i}s_\omega\nu)\right] e^{-\mathrm{i}s_\omega\nu(\pm\psi-\frac{1}{2}\pi)}. \tag{D14}$$

$$\delta_x\dot{\delta}_y\delta_t E_z = -E_0(-\omega)\left[\frac{k_x k_y-\mathrm{i}s_\omega(k^2_y+k^2_z)}{k^3_z}A + \frac{(k_x-\mathrm{i}s_\omega k_y)k_y}{k_z}B - \frac{(k_x-\mathrm{i}s_\omega k_y)k_x}{k_z k^2_\perp}A(\pm \mathrm{i}s_\omega\nu)\right] e^{-\mathrm{i}s_\omega\nu(\pm\psi-\frac{1}{2}\pi)}. \tag{D15}$$

Upon substituting Eqs.(D10)-(D15) into Eqs.(23), the first contributions $\mathcal{L}^{(1)}_x$, $\mathcal{L}^{(1)}_y$, $\mathcal{L}^{(1)}_z$ to the components of $\boldsymbol{\mathcal{L}}$ are found to be

$$\begin{aligned}
\mathcal{L}^{(1)}_x &= \frac{\mathrm{i}}{2\pi Z_0\alpha^{2(\nu+1)}}\int_{-\infty}^{\infty}\omega^{-1}|E(\omega)|^2\left(\frac{ck_z}{\omega}\right)\left\{\left[-Bk_y + A(\pm \mathrm{i}s_\omega\nu)\frac{k_x}{k^2_\perp}\right]\left[\frac{k_x(k_x+\mathrm{i}s_\omega k_y)}{k_z}+k_z\right]A\right.\\
&\qquad\left. +\mathrm{i}s_\omega\left[Bk_y - A(\pm \mathrm{i}s_\omega\nu)\frac{k_x}{k^2_\perp}\right]\left[\frac{k_y(k_x+\mathrm{i}s_\omega k_y)}{k_z}+\mathrm{i}s_\omega k_z\right]A\right\}\mathrm{d}k_x\mathrm{d}k_y\mathrm{d}\omega\\
&= \frac{\mathrm{i}}{2\pi Z_0\alpha^{2(\nu+1)}}\int_{-\infty}^{\infty}\omega^{-1}|E(\omega)|^2(k^2_\perp+2k^2_z)\left[A^2(\pm \mathrm{i}s_\omega\nu)\frac{(ck_x/\omega)}{k^2_\perp} - AB\left(\frac{ck_y}{\omega}\right)\right]\mathrm{d}k_x\mathrm{d}k_y\mathrm{d}\omega.
\end{aligned} \tag{D16}$$

$$\begin{aligned}
\mathcal{L}^{(1)}_y &= \frac{1}{\mathrm{i}2\pi Z_0\alpha^{2(\nu+1)}}\int_{-\infty}^{\infty}\omega^{-1}|E(\omega)|^2\left(\frac{ck_z}{\omega}\right)\left\{\mathrm{i}s_\omega\left[Bk_x + A(\pm \mathrm{i}s_\omega\nu)\frac{k_y}{k^2_\perp}\right]\left[\frac{k_y(k_x+\mathrm{i}s_\omega k_y)}{k_z}+\mathrm{i}s_\omega k_z\right]A\right.\\
&\qquad\left. -\left[Bk_x + A(\pm \mathrm{i}s_\omega\nu)\frac{k_y}{k^2_\perp}\right]\left[\frac{k_x(k_x+\mathrm{i}s_\omega k_y)}{k_z}+k_z\right]A\right\}\mathrm{d}k_x\mathrm{d}k_y\mathrm{d}\omega\\
&= \frac{\mathrm{i}}{2\pi Z_0\alpha^{2(\nu+1)}}\int_{-\infty}^{\infty}\omega^{-1}|E(\omega)|^2(k^2_\perp+2k^2_z)\left[A^2(\pm \mathrm{i}s_\omega\nu)\frac{(ck_y/\omega)}{k^2_\perp} + AB\left(\frac{ck_x}{\omega}\right)\right]\mathrm{d}k_x\mathrm{d}k_y\mathrm{d}\omega.
\end{aligned} \tag{D17}$$

$$\mathcal{L}_z^{(1)} = \frac{1}{\mathrm{i}2\pi Z_0 \alpha^{2(\nu+1)}} \int_{-\infty}^{\infty} \omega^{-1} |E(\omega)|^2 \left(\frac{ck_z}{\omega}\right) \times$$

$$\left\{ \left[ \frac{(k_x^2+k_z^2) - \mathrm{i}s_\omega k_x k_y}{k_z^3} A^2 + \frac{(k_x - \mathrm{i}s_\omega k_y)k_x}{k_z} AB + \frac{(k_x - \mathrm{i}s_\omega k_y)k_y}{k_z k_\perp^2} A^2 (\pm \mathrm{i}s_\omega \nu) \right] \left[ \left(\frac{k_y}{k_z}\right)(k_x + \mathrm{i}s_\omega k_y) + \mathrm{i}s_\omega k_z \right] \right.$$

$$- \left[ \frac{k_x k_y - \mathrm{i}s_\omega (k_y^2 + k_z^2)}{k_z^3} A^2 + \frac{(k_x - \mathrm{i}s_\omega k_y)k_y}{k_z} AB - \frac{(k_x - \mathrm{i}s_\omega k_y)k_x}{k_z k_\perp^2} A^2 (\pm \mathrm{i}s_\omega \nu) \right] \left[ \left(\frac{k_x}{k_z}\right)(k_x + \mathrm{i}s_\omega k_y) + k_z \right]$$

$$\left. + \left( ABk_x + A^2(\pm \mathrm{i}s_\omega \nu) \frac{k_y}{k_\perp^2} - \mathrm{i}s_\omega \left[ ABk_y - A^2 (\pm \mathrm{i}s_\omega \nu) \frac{k_x}{k_\perp^2} \right] \right) (k_y - \mathrm{i}s_\omega k_x) \right\} \mathrm{d}k_x \mathrm{d}k_y \mathrm{d}\omega. \quad \text{(D18)}$$

The odd symmetry of the integrands of $\mathcal{L}_x^{(1)}$ and $\mathcal{L}_y^{(1)}$ with respect to the origin of the $k_x k_y$-plane ensures that $\mathcal{L}_x^{(1)} = \mathcal{L}_y^{(1)} = 0$. As for the $z$-component of $\boldsymbol{\mathcal{L}}$, straightforward algebraic manipulations lead to the following compact expression for $\mathcal{L}_z^{(1)}$ of Eq.(D18):

$$\mathcal{L}_z^{(1)} = \frac{1}{2\pi Z_0 \alpha^{2(\nu+1)}} \int_{-\infty}^{\infty} s_\omega \omega^{-1} |E(\omega)|^2 \left[ \frac{2\omega}{ck_z} \pm \left( \frac{ck_z}{\omega} + \frac{\omega}{ck_z} \right) \nu \right] k_\perp^{2\nu} [L_n^\nu (k_\perp^2/\alpha)]^2 e^{-k_\perp^2/\alpha} \mathrm{d}k_x \mathrm{d}k_y \mathrm{d}\omega. \quad \text{(D19)}$$

Note that the $\pm$ signs pertain to counterclockwise and clockwise vortices, whose phase modulation factors are $e^{\pm \mathrm{i}\nu\varphi}$, and that $s_\omega \omega^{-1}$ can equivalently be written as $|\omega|^{-1}$. The remaining contributions $\mathcal{L}_x^{(2)}, \mathcal{L}_y^{(2)}, \mathcal{L}_z^{(2)}$ to the components of $\boldsymbol{\mathcal{L}}$ come from integrals of the form

$$\iiint_{-\infty}^{\infty} \delta(k_x + k_x')\delta(k_y + k_y')\dot{\delta}(\omega + \omega') E_x(\boldsymbol{k}', \omega') k_\perp'^{\nu} L_n^\nu (k_\perp'^2/\alpha) e^{-k_\perp'^2/2\alpha} e^{\mathrm{i}s_\omega' \nu(\pm\psi' - \frac{1}{2}\pi)} \mathrm{d}k_x' \mathrm{d}k_y' \mathrm{d}\omega'. \quad \text{(D20)}$$

In some instances, the integrand in Eq.(D20) will have a $k_x/k_z'$ or $k_y/k_z'$ coefficient as well; see Eqs.(24)-(26). Thus, the fifth and final step of computing the components of $\boldsymbol{\mathcal{L}}$ involves evaluating such triple integrals. Here, one substitutes $(-k_x, -k_y)$ for $(k_x', k_y')$, differentiates with respect to $\omega'$, substitutes $-\omega$ for $\omega'$ and, finally, flips the sign of the entire expression. Carrying out these operations for the individual terms that appear in Eqs.(23), we find

$$(k_y/k_z')\delta_x\delta_y\dot{\delta}_t E_x' H_y = \frac{cE_0(\omega)}{\omega Z_0} \left[ \dot{E}_0(-\omega) \left(\frac{k_y}{k_z}\right) + E_0(-\omega) \left(\frac{c^2 k_z}{\omega}\right)^{-1} \left(\frac{k_y}{k_z^2}\right) \right] \left[ \frac{k_x(k_x + \mathrm{i}s_\omega k_y)}{k_z} + k_z \right] A^2. \quad \text{(D21)}$$

$$\delta_x\delta_y\dot{\delta}_t E_x' H_z = \frac{cE_0(\omega)\dot{E}_0(-\omega)}{\omega Z_0} (k_y - \mathrm{i}s_\omega k_x) A^2. \quad \text{(D22)}$$

$$(k_y/k_z')\delta_x\delta_y\dot{\delta}_t E_y' H_x = -\frac{cE_0(\omega)}{\omega Z_0} (-\mathrm{i}s_\omega) \left[ \dot{E}_0(-\omega) \left(\frac{k_y}{k_z}\right) + E_0(-\omega) \left(\frac{c^2 k_z}{\omega}\right)^{-1} \left(\frac{k_y}{k_z^2}\right) \right] \left[ \frac{k_y(k_x + \mathrm{i}s_\omega k_y)}{k_z} + \mathrm{i}s_\omega k_z \right] A^2. \quad \text{(D23)}$$

$$\delta_x\delta_y\dot{\delta}_t E_z' H_x = -\frac{cE_0(\omega)}{\omega Z_0} \left[ \dot{E}_0(-\omega) \left(\frac{k_x - \mathrm{i}s_\omega k_y}{k_z}\right) + E_0(-\omega) \left(\frac{c^2 k_z}{\omega}\right)^{-1} \left(\frac{k_x - \mathrm{i}s_\omega k_y}{k_z^2}\right) \right] \left[ \frac{k_y(k_x + \mathrm{i}s_\omega k_y)}{k_z} + \mathrm{i}s_\omega k_z \right] A^2. \quad \text{(D24)}$$

$$\delta_x\delta_y\dot{\delta}_t E_y' H_z = \frac{cE_0(\omega)\dot{E}_0(-\omega)}{\omega Z_0} (-\mathrm{i}s_\omega)(k_y - \mathrm{i}s_\omega k_x) A^2. \quad \text{(D25)}$$

$$(k_x/k_z')\delta_x\delta_y\dot{\delta}_t E_y' H_x = -\frac{cE_0(\omega)}{\omega Z_0} (-\mathrm{i}s_\omega) \left[ \left(\frac{k_x}{k_z}\right) \dot{E}_0(-\omega) + \left(\frac{c^2 k_z}{\omega}\right)^{-1} \left(\frac{k_x}{k_z^2}\right) E_0(-\omega) \right] \left[ \frac{k_y(k_x + \mathrm{i}s_\omega k_y)}{k_z} + \mathrm{i}s_\omega k_z \right] A^2. \quad \text{(D26)}$$

$$\delta_x\delta_y\dot{\delta}_t E_z' H_y = \frac{cE_0(\omega)}{\omega Z_0} \left[ \dot{E}_0(-\omega) \left(\frac{k_x - \mathrm{i}s_\omega k_y}{k_z}\right) + E_0(-\omega) \left(\frac{c^2 k_z}{\omega}\right)^{-1} \left(\frac{k_x - \mathrm{i}s_\omega k_y}{k_z^2}\right) \right] \left[ \frac{k_x(k_x + \mathrm{i}s_\omega k_y)}{k_z} + k_z \right] A^2. \quad \text{(D27)}$$

$$(k_x/k_z')\delta_x\delta_y\dot{\delta}_t E_x' H_y = \frac{cE_0(\omega)}{\omega Z_0}\left[\left(\frac{k_x}{k_z}\right)\dot{E}_0(-\omega) + \left(\frac{c^2 k_z}{\omega}\right)^{-1}\left(\frac{k_x}{k_z^2}\right)E_0(-\omega)\right]\left[\frac{k_x(k_x+\mathrm{i}s_\omega k_y)}{k_z} + k_z\right]A^2. \quad \text{(D28)}$$

$$(k_x/k_z')\delta_x\delta_y\dot{\delta}_t E_z' H_x = \frac{cE_0(\omega)}{\omega Z_0}\left[\dot{E}_0(-\omega)\frac{k_x(k_x-\mathrm{i}s_\omega k_y)}{k_z^2} + 2E_0(-\omega)\left(\frac{c^2 k_z}{\omega}\right)^{-1}\frac{k_x(k_x-\mathrm{i}s_\omega k_y)}{k_z^3}\right]\left[\frac{k_y(k_x+\mathrm{i}s_\omega k_y)}{k_z} + \mathrm{i}s_\omega k_z\right]A^2. \quad \text{(D29)}$$

$$(k_y/k_z')\delta_x\delta_y\dot{\delta}_t E_z' H_y = -\frac{cE_0(\omega)}{\omega Z_0}\left[\dot{E}_0(-\omega)\frac{k_y(k_x-\mathrm{i}s_\omega k_y)}{k_z^2} + 2\left(\frac{c^2 k_z}{\omega}\right)^{-1}E_0(-\omega)\frac{k_y(k_x-\mathrm{i}s_\omega k_y)}{k_z^3}\right]\left[\frac{k_x(k_x+\mathrm{i}s_\omega k_y)}{k_z} + k_z\right]A^2. \quad \text{(D30)}$$

$$(k_x/k_z')\delta_x\delta_y\dot{\delta}_t E_x' H_z = -\frac{cE_0(\omega)}{\omega Z_0}\left[\left(\frac{k_x}{k_z}\right)\dot{E}_0(-\omega) + \left(\frac{c^2 k_z}{\omega}\right)^{-1}\left(\frac{k_x}{k_z^2}\right)E_0(-\omega)\right](k_y - \mathrm{i}s_\omega k_x)A^2. \quad \text{(D31)}$$

$$(k_y/k_z')\delta_x\delta_y\dot{\delta}_t E_y' H_z = -\frac{cE_0(\omega)}{\omega Z_0}(-\mathrm{i}s_\omega)\left[\left(\frac{k_y}{k_z}\right)\dot{E}_0(-\omega) + \left(\frac{c^2 k_z}{\omega}\right)^{-1}\left(\frac{k_y}{k_z^2}\right)E_0(-\omega)\right](k_y - \mathrm{i}s_\omega k_x)A^2. \quad \text{(D32)}$$

The relevant combinations of the above terms for $\mathcal{L}_x$, $\mathcal{L}_y$, and $\mathcal{L}_z$ are found to be

$$\left(\frac{c^2 k_z}{\omega}\right)^2\left[-\left(\frac{k_y}{k_z'}\right)\delta_x\delta_y\dot{\delta}_t E_x' H_y + \delta_x\delta_y\dot{\delta}_t E_x' H_z + \left(\frac{k_y}{k_z'}\right)\delta_x\delta_y\dot{\delta}_t E_y' H_x - \delta_x\delta_y\dot{\delta}_t E_z' H_x\right] = \frac{c^3|E_0(\omega)|^2}{Z_0\omega^2}(\mathrm{i}s_\omega k_x - k_y)A^2. \quad \text{(D33)}$$

$$\left(\frac{c^2 k_z}{\omega}\right)^2\left[\delta_x\delta_y\dot{\delta}_t E_y' H_z - \left(\frac{k_x}{k_z'}\right)\delta_x\delta_y\dot{\delta}_t E_y' H_x - \delta_x\delta_y\dot{\delta}_t E_z' H_y + \left(\frac{k_x}{k_z'}\right)\delta_x\delta_y\dot{\delta}_t E_x' H_y\right] = \frac{c^3|E_0(\omega)|^2}{Z_0\omega^2}(k_x + \mathrm{i}s_\omega k_y)A^2. \quad \text{(D34)}$$

$$\left(\frac{c^2 k_z}{\omega}\right)^2\left[-\left(\frac{k_x}{k_z'}\right)\delta_x\delta_y\dot{\delta}_t E_z' H_x - \left(\frac{k_y}{k_z'}\right)\delta_x\delta_y\dot{\delta}_t E_z' H_y + \left(\frac{k_x}{k_z'}\right)\delta_x\delta_y\dot{\delta}_t E_x' H_z + \left(\frac{k_y}{k_z'}\right)\delta_x\delta_y\dot{\delta}_t E_y' H_z\right]$$
$$= -\mathrm{i}s_\omega\frac{c^2|E_0(\omega)|^2}{Z_0\omega}\left(\frac{ck_z}{\omega}\right)\left(\frac{k_\perp}{k_z}\right)^2 A^2. \quad \text{(D35)}$$

Once again, the odd symmetry (with respect to the origin of the $k_x k_y$-plane) of the first two of the above expressions ensures that their net contributions to $\mathcal{L}_x$ and $\mathcal{L}_y$ would vanish. The expression in Eq.(D35), however, makes a small but important contribution to $\mathcal{L}_z$, which has to be taken into account. As before, $s_\omega\omega^{-1} = |\omega|^{-1}$, and the contribution of the expression in Eq.(D35) to the wave-packet's angular momentum along the $z$-axis is given by

$$\mathcal{L}_z^{(2)} = -\frac{1}{Z_0\alpha^{2(\nu+1)}}\int_{\omega=-\infty}^{\infty}\frac{|E_0(\omega)|^2}{|\omega|}\int_{k_\perp=0}^{\infty}(ck_z/\omega)(k_\perp/k_z)^2 k_\perp^{2\nu}[L_n^\nu(k_\perp^2/\alpha)]^2 e^{-k_\perp^2/\alpha}k_\perp \mathrm{d}k_\perp \mathrm{d}\omega. \quad \text{(D36)}$$

Combining Eqs.(D19) and (D36), we finally arrive at a formula for the packet's total angular momentum along its propagation direction; that is,

$$\mathcal{L}_z = \mathcal{L}_z^{(1)} + \mathcal{L}_z^{(2)} = \frac{1\pm\nu}{Z_0\alpha^{2(\nu+1)}}\int_{\omega=-\infty}^{\infty}\frac{|E_0(\omega)|^2}{|\omega|}\int_{k_\perp=0}^{\infty}\overbrace{\left(\frac{ck_z}{\omega} + \frac{\omega}{ck_z}\right)}^{\cong 2}k_\perp^{2\nu}[L_n^\nu(k_\perp^2/\alpha)]^2 e^{-k_\perp^2/\alpha}k_\perp \mathrm{d}k_\perp \mathrm{d}\omega. \quad \text{(D37)}$$

Assuming the obliquity factor $ck_z/\omega$ is not too far below 1, the integral over the $k_x k_y$-plane yields $\alpha^{\nu+1}(n+\nu)!/n!$, while the integral of $|E_0(\omega)|^2$ over $\omega$ gives $2\pi\int_{-\infty}^{\infty}E^2(t)\mathrm{d}t$. For a narrowband wave-packet centered at the frequency $\omega$, the overall $\mathcal{L}_z$ thus turns out to be nearly equal to $(1\pm\nu)\mathcal{E}/\omega$, with $\mathcal{E}$ being the total EM energy of the wave-packet; see Eq.(17).

## Appendix E

We present a brief description of the Bessel-Gauss beams, which have certain similarities but also substantial differences with the class of Laguerre-Gauss beams. The relevant integrals pertaining to the Fourier transform of Bessel-Gauss beams in two-dimensional space are

> $I_\nu(x) = \mathrm{i}^{-\nu} J_\nu(\mathrm{i}x)$ is a modified Bessel function of the first kind, order $\nu$.

$$\int_0^\infty e^{-\zeta x^2/2} J_\nu(\eta x) J_\nu(\kappa x)\, x\,\mathrm{d}x = \zeta^{-1} e^{-(\eta^2+\kappa^2)/2\zeta} I_\nu(\eta\kappa/\zeta). \quad \text{(G\&R 6.633-2)} \qquad \text{(E1)}$$

$$\int_0^\infty e^{-\zeta\kappa^2/2} I_\nu(\eta\kappa) J_\nu(\kappa x)\, \kappa\,\mathrm{d}\kappa = \zeta^{-1} e^{(\eta^2-x^2)/2\zeta} J_\nu(\eta x/\zeta). \quad \text{(G\&R 6.633-4)} \qquad \text{(E2)}$$

The scalar amplitude profile of a Bessel-Gauss beam in the $xy$-plane at $z = 0$ is defined as

$$a(x, y, z = 0) = a(r, \varphi, z = 0) = a_0 J_\nu(r/r_0) e^{-(r/w_0)^2} e^{\pm \mathrm{i}\nu\varphi}. \qquad \text{(E3)}$$

Here, the amplitude $a_0$ is a (generally complex) constant; $J_\nu(\cdot)$ is a Bessel function of the first kind, integer order $\nu \geq 0$; the parameters $r_0 > 0$ and $w_0 > 0$ are real-valued constants; and the $\pm$ signs indicate the sense of the beam's vorticity as either clockwise or counterclockwise. The beam's integrated intensity is given by

$$\int_0^\infty |a_0|^2 J_\nu^2(r/r_0) e^{-2(r/w_0)^2} 2\pi r\mathrm{d}r = ½\pi |a_0|^2 w_0^2 e^{-(w_0/2r_0)^2} I_\nu(w_0^2/4r_0^2). \qquad \text{(E4)}$$

The Fourier transform of the amplitude profile is readily computed with the aid of Eq.(E1), as follows:

$$\iint_{-\infty}^{\infty} a_0 J_\nu(r/r_0) e^{-(r/w_0)^2} e^{\pm \mathrm{i}\nu\varphi} e^{-\mathrm{i}(k_x x + k_y y)} \mathrm{d}x\mathrm{d}y$$

$$= a_0 \int_{r=0}^{\infty} J_\nu(r/r_0) e^{-(r/w_0)^2} \left[\int_{\varphi=0}^{2\pi} e^{\pm \mathrm{i}\nu\varphi} e^{-\mathrm{i}k_\perp r\cos(\psi-\varphi)} \mathrm{d}\varphi\right] r\mathrm{d}r$$

$$= a_0 \int_{r=0}^{\infty} J_\nu(r/r_0) e^{-(r/w_0)^2} \left[e^{\mathrm{i}\nu(\pm\psi - ½\pi)} \int_{\theta=0}^{2\pi} e^{-\mathrm{i}\nu\theta} e^{\mathrm{i}k_\perp r\sin\theta} \mathrm{d}\theta\right] r\mathrm{d}r$$

$$= 2\pi a_0 e^{\mathrm{i}\nu(\pm\psi - ½\pi)} \int_0^\infty J_\nu(r/r_0) J_\nu(k_\perp r) e^{-(r/w_0)^2} r\mathrm{d}r$$

$$= \pi a_0 w_0^2 e^{\mathrm{i}\nu(\pm\psi - ½\pi)} e^{-w_0^2(k_\perp^2 + r_0^{-2})/4} I_\nu\left(\frac{w_0^2 k_\perp}{2r_0}\right). \qquad \text{(E5)}$$

The inverse Fourier transform identity may now be confirmed with the aid of Eq.(E2); that is,

$$(2\pi)^{-2} \iint_{-\infty}^{\infty} \pi a_0 w_0^2 e^{\mathrm{i}\nu(\pm\psi - ½\pi)} e^{-w_0^2(k_\perp^2 + r_0^{-2})/4} I_\nu\left(\frac{w_0^2 k_\perp}{2r_0}\right) e^{\mathrm{i}(k_x x + k_y y)} \mathrm{d}k_x \mathrm{d}k_y$$

$$= \frac{a_0 w_0^2 e^{-(w_0/2r_0)^2}}{4\pi} \int_{k_\perp=0}^{\infty} e^{-(w_0 k_\perp/2)^2} I_\nu\left(\frac{w_0^2 k_\perp}{2r_0}\right) \left[\int_{\psi=0}^{2\pi} e^{\mathrm{i}\nu(\pm\psi - ½\pi)} e^{\mathrm{i}k_\perp r\cos(\varphi-\psi)} \mathrm{d}\psi\right] k_\perp \mathrm{d}k_\perp$$

$$= ½ a_0 w_0^2 e^{-(w_0/2r_0)^2} e^{\pm \mathrm{i}\nu\varphi} \int_0^\infty e^{-(w_0 k_\perp/2)^2} I_\nu\left(\frac{w_0^2 k_\perp}{2r_0}\right) J_\nu(k_\perp r) k_\perp \mathrm{d}k_\perp$$

$$= a_0 J_\nu(r/r_0) e^{-(r/w_0)^2} e^{\pm \mathrm{i}\nu\varphi}. \qquad \text{(E6)}$$

Upon propagation in free space, the beam profile (in the paraxial approximation) becomes

$$a(x, y, z) = (2\pi)^{-2} \iint_{-\infty}^{\infty} \pi a_0 w_0^2 e^{\mathrm{i}\nu(\pm\psi - ½\pi)} e^{-w_0^2(k_\perp^2 + r_0^{-2})/4} I_\nu\left(\frac{w_0^2 k_\perp}{2r_0}\right) e^{\mathrm{i}(k_x x + k_y y + k_z z)} \mathrm{d}k_x \mathrm{d}k_y$$

$$\cong \tfrac{1}{2} a_0 w_0^2 e^{-(w_0/2r_0)^2} e^{\pm \mathrm{i}\nu\varphi} \int_{k_\perp=0}^{\infty} e^{-(w_0 k_\perp/2)^2} I_\nu\left(\frac{w_0^2 k_\perp}{2r_0}\right) J_\nu(k_\perp r) e^{\mathrm{i}k_0[1-\frac{1}{2}(k_\perp/k_0)^2]z} k_\perp \mathrm{d}k_\perp$$

$$= \tfrac{1}{2} a_0 w_0^2 e^{-(w_0/2r_0)^2} e^{\pm \mathrm{i}\nu\varphi} e^{\mathrm{i}k_0 z} \int_0^{\infty} e^{-[w_0^2+\mathrm{i}(2z/k_0)]k_\perp^2/4} I_\nu\left(\frac{w_0^2 k_\perp}{2r_0}\right) J_\nu(k_\perp r) k_\perp \mathrm{d}k_\perp$$

$$= \frac{a_0 \exp(\mathrm{i}k_0 z)}{1+\mathrm{i}(2z/k_0 w_0^2)} J_\nu\left[\frac{r/r_0}{1+\mathrm{i}(2z/k_0 w_0^2)}\right] \exp\left[-\frac{(r/w_0)^2 + \mathrm{i}(z/2k_0 r_0^2)}{1+\mathrm{i}(2z/k_0 w_0^2)}\right] e^{\pm \mathrm{i}\nu\varphi}. \qquad \text{(E7)}$$

Note that, at $z \neq 0$, the argument of the Bessel function is complex-valued, indicating that the beam profile is no longer a scaled version of what it was at $z = 0$. This is a main difference between Bessel-Gauss and Laguerre-Gauss beams, where the latter retain the general beam profile at all values of $z$. Nevertheless, so long as $k_0 z \ll (k_0 w_0)^2$, the argument of $J_\nu(\cdot)$ in Eq.(E7) remains, to good approximation, nearly the same as $r/r_0$, indicating that, for short propagation distances, the essential diffraction-free nature of pure Bessel beams is preserved. In the far field, however, the Bessel function's argument approaches a purely imaginary value, turning the cross-sectional profile of the beam into $I_\nu(k_0 w_0^2 r/2r_0 z)$, which, in combination with the Gaussian envelope, produces a bright ring of light in the far field.

Note that the exponential factor in Eq.(E7) has a real part, which is associated with the Gaussian envelope of the beam, and an imaginary part, which produces the wavefront curvature. When separated into its real and imaginary parts, the exponential factor in Eq.(E7) becomes

$$\exp\left[-\frac{r^2 + (z/k_0 r_0)^2}{w_0^2 + (2z/k_0 w_0)^2}\right] \exp\left[\mathrm{i}\frac{(r/w_0)^2 - (w_0/2r_0)^2}{(k_0 w_0^2/2z) + (2z/k_0 w_0^2)}\right]. \qquad \text{(E8)}$$

Aside from the curvature phase-factor and the intrinsic phase associated with a Bessel function of complex argument, the only other $z$-dependent phase-factor in Eq.(E7) derives from the leading coefficient, namely, $\exp\{\mathrm{i}[k_0 z - \tan^{-1}(2z/k_0 w_0^2)]\}$.

## References

1. L. Allen, S. M. Barnett, and M. J. Padgett, *Optical Angular Momentum*, Institute of Physics Publishing, Bristol, UK (2003).
2. L. Allen, M. W. Beijersbergen, R. J. C. Spreeuw, and J. P. Woerdman, "Orbital angular momentum of light and the transformation of Laguerre-Gaussian laser modes," *Physical Review A* **45**, 8185-89 (1992).
3. M. W. Beijersbergen, L. Allen, H. E. L. O. van der Veen, and J. P. Woerdman, "Astigmatic laser mode converters and transfer of orbital angular momentum," *Optics Communications* **96**, 123-132 (1993).
4. S. M. Barnett and L. Allen, "Orbital angular momentum and nonparaxial light beams," *Optics Communications* **110**, 670-678 (1994).
5. L. Marrucci, E. Karimi, S. Slussarenko, B. Piccirillo, E. Santamato, E. Nagali, and F. Sciarrino, "Spin-to-orbital conversion of the angular momentum of light and its classical and quantum applications," *J. Opt.* **13**, 064001 (2011).
6. F. Gori, G. Guattari, and C. Padovani, "Bessel-Gauss Beams," *Optics Communications* **64**, 491-495 (1987).
7. M. Born and E. Wolf, *Principles of Optics* (7th edition), Cambridge University Press, Cambridge, UK (2002).
8. A. E. Siegman, *Lasers*, University Science Books, Sausalito, California (1986).
9. I. S. Gradshteyn and I. M. Ryzhik, *Table of Integrals, Series, and Products*, 7th edition, Academic Press, New York (2007).
10. J. D. Jackson, *Classical Electrodynamics* (3rd edition), Wiley, New York (1999).
11. M. Mansuripur, *Field, Force, Energy and Momentum in Classical Electrodynamics* (revised edition), Bentham Science Publishers, Sharjah, UAE (2017).
12. T. A. Nieminen, A. B. Stilgoe, N. R. Heckenberg, and H. Rubinsztein-Dunlop, "Angular momentum of a strongly focused Gaussian beam," *Journal of Optics A: Pure and Applied Optics* **10**, 115005 (2008).